\documentclass{PoS}

\usepackage{cite}
\title{Origin of the p-Nuclei in Explosive Nucleosynthesis}

\ShortTitle{Origin of the p-Nuclei}

\author{\speaker{Thomas Rauscher}%
         \\
        Department of Physics\\
        University of Basel\\
        E-mail: \email{Thomas.Rauscher@unibas.ch}}

\abstract{A number of naturally occurring, proton-rich nuclides (the p-nuclei) cannot be made in the s- and r-process.
It has been found that massive
stars can produce p-nuclei through photodisintegration of pre-existing intermediate and heavy nuclei. This
so-called $\gamma$-process requires sufficiently high temperatures and occurs in pre-explosive or explosive O/Ne
burning, depending on the mass of the star. Although the $\gamma$-process has been successful in producing a large
range of p-nuclei, two mass regions remain problematic, $A<110$ and $150<A<165$, where a number of p-nuclei are severely
underproduced. The origin of the problems is yet to be identified. A large number of unstable nuclei with only
theoretically predicted reaction rates are included in the reaction network and thus the nuclear input may
involve uncertainties. Deficiencies in charged-particle optical potentials at $\gamma$-process temperatures have
been found for nuclei at stability. On the other hand, the $\gamma$-process conditions (temperature profiles, entropy
of the O shell, seed composition) also sensitively depend on details of the stellar structure and evolution, as
well as on the initial metallicity. Nevertheless, especially the deficient low-mass p-nuclei may call for an
additional production process or site, such as production in (subChandrasekhar) type Ia
supernovae. Also the (im)possibility of a synthesis in the rp- and $\nu$p-processes is discussed.
Were this the case, the production of p-nuclei would be realized as a superposition of several different
processes, not necessarily in the same site.}

\FullConference{11th Symposium on Nuclei in the Cosmos\\
		 19-23 July 2010 \\
		 Heidelberg, Germany.}

\begin{document}

\section{The case of the missing nuclides}

\noindent
When B$^2$FH \cite{b2fh} and A. G. W. Cameron \cite{agw}
postulated the s- and r-processes for the production of intermediate and heavy nuclei beyond iron, they realized that
a number of proton-rich isotopes can never be synthesized through sequences of only neutron captures and $\beta^-$ decays. This
required the postulation of a third process. It was termed \textit{p-process} because it was thought to proceed via proton captures
at high temperature, perhaps even reaching (partial) (p,$\gamma$)-($\gamma$,p) equilibrium. B$^2$FH tentatively placed this
nucleosynthesis process in the H-rich envelope of type II supernovae but it was later realized that the required temperatures
are not acquired there \cite{autru,arngorp}. This also shed doubts on the feasibility to use proton captures for producing \textit{all}
of the nuclides missing from the s- and r-process production.

It is somewhat confusing that in the
literature the name "p-process" is sometimes used for a proton capture process in the spirit of B$^2$FH but also sometimes
taken as a token subsuming whatever production mechanism(s) is/are found to be
responsible for the p-nuclides. For an easier distinction of the production processes, perhaps
modern nomenclature better focuses on naming the nuclides in question
the \textit{p-nuclides} (they were called ``excluded isotopes'' by \cite{agw}) and use appropriate names for the processes involved.

Compared to the bulk of natural isotopes, the p-nuclei generally show abundances which are $1-2$ orders of magnitude lower.
Historically there were 35 p-nuclides identified, with $^{74}$Se being the lightest and $^{196}$Hg the heaviest. It is to be
noted, however, that this assignment depends on the state-of-the-art of the s-process models (just like the "observed"
r-abundances depend on them) and also on estimates of r-process contributions (e.g., to $^{113}$In and $^{115}$Sn) \cite{dill30,nemeth}.
Already B$^2$FH realized that, e.g., the abundance of $^{164}$Er stands out among its neighboring p-nuclides and may thus contain
considerable contributions from the s-process. It was indeed found that there are large s-process contributions to $^{164}$Er and
also to $^{152}$Gd \cite{arl99}, thus possibly removing them from the list of p-isotopes. If also the abundances of
$^{113}$In and $^{115}$Sn can be explained by modifications of the s-process and/or contributions from the r-process \cite{nemeth},
this would leave only 31 p-isotopes to be explained by other processes.

A special obstacle not found in the investigation of the s- and r-process is the fact that there are no elements dominated by p-isotopes. Therefore,
our knowledge of p-abundances are limited to solar system abundances derived from meteoritic material and terrestrial isotopic
compositions. Before astronomical observations of isotopic abundances at the required discrimination level become feasible (if ever),
it is impossible to determine p-abundances in stars of different metallicities and thus to obtain the galactic chemical evolution
picture directly. On the other hand, depending on the actual p-production mechanism this may also be problematic for determining early
s-process contributions. If the p-nuclides (or some of them) turn out to be primary (i.e., independent of metallicity) or have a different dependence of production on metallicity than the s-process (perhaps by initially originating from r-nuclei),
they may give a larger contribution to elemental abundances in old stars than the s-process.

\section{Processes and sites possibly contributing to the production of p-nuclei}

\noindent
There are several possibilities to get to the proton-rich side. Sequences of proton captures may reach a p-isotope from elements
with lower charge number. They are suppressed by the Coulomb barriers, however, and it is not possible to arbitrarily compensate
for that by just requiring higher plasma temperatures. At high temperature ($\gamma$,p) reactions become faster than proton captures
and prevent the build-up of proton-rich nuclides. Photodisintegrations are an alternative way to make p-nuclei, either by directly
producing them through destruction of their neutron-richer isotopes through sequences of ($\gamma$,n) reactions (these are the
predominant photodisintegration processes for most stable nuclei), or by flows from
heavier elements via ($\gamma$,p) and ($\gamma$,$\alpha$) reactions and $\beta$-decays. The latter possibility requires higher
temperatures because unstable, proton-rich nuclei have to be produced through sequences of ($\gamma$,n) reactions first.

There are three ingredients influencing the resulting p-abundances and these three are differently combined in the various sites
proposed as the birthplace of the p-nuclides. The first one is, obviously, the temperature variation as a function of time, defining
the timescale of the process and the peak temperature. This already points us to explosive conditions which accommodate both the
necessary temperatures for photodisintegrations (or proton captures on highly charged nuclei) and a short timescale. The latter is
required because it has to be avoided that too much material is transformed in order to achieve the tiny solar relative p-abundances
(assuming that these are typical). The second parameter is the proton density. While photodisintegrations and $\beta$ decays are
not sensitive to the proton abundance, proton captures are. With a high number of protons available, proton captures can prevail
over ($\gamma$,p) reactions even at high temperature. Last but nor least, we can vary the seed abundances, i.e. the number and
composition of nuclei on which the photodisintegrations or proton captures act initially. If the final p-abundances depend
sensitively on these seeds (and they do in most suggested sites), this implies that the p-nuclei are secondary and thus depending
on some s- and/or r-process nuclides being already present in the material.

Here, only the most relevant suggested sites and processes for the production of p-nuclides can be mentioned (for further details and
additional alternatives, see \cite{arngorp} and references therein). The most favored process so far
is the \textit{$\gamma$-process} occurring during explosive O/Ne-shell burning in massive stars \cite{woohow,arn,rayet95,rhhw02}. It was realized early
that the abundances of most p-nuclei are inversely correlated with their photodisintegration rates \cite{agw,woohow}, pointing to an important
contribution of photodisintegration. Massive stars provide the required conditions of transforming s- and r-process material already present
in the protostellar cloud. During the final core-collapse supernova a shockwave ejects and heats the outer layers of the star, in just the right amount needed to produce p-nuclei through photodisintegration. (Stars with higher mass may reach $\gamma$-process
temperatures of $2\leq T_9 \leq 3$ already pre-explosively \cite{rhhw02} although some of those pre-explosive p-nuclei may be destroyed again in the explosion.)
The $\gamma$-process occurs naturally in massive stars and does not require any artifical fine-tuning. It is appealing because it can co-produce
the bulk of p-nuclides within a single site. All p-nuclei are secondary. The very rare $^{138}$La can be produced through neutrino reactions with neutrinos emitted
by the nascent neutron star emerging from the core collapse (the $\nu$-process) \cite{neutrino}. The equally rare $^{180m}$Ta probably
also received a large contribution from the $\nu$-process. Nevertheless, there are longstanding shortcomings in the production of certain mass ranges which persist also in current models. The light p-nuclei with $A<110$ are strongly underproduced and also the mass range $150\leq A \leq165$ seems problematic.
This has triggered a number of investigations in astrophysics and nuclear physics aimed at resolving these deficiences.

An interesting alternative is the possible production of p-nuclei in hot, proton-rich environments. At least the light p-nuclides
could then be made by proton captures. In combination with a modified seed abundance (strong enrichment in s-process nuclei), again
the whole range of p-nuclei could be synthesized possibly within one appropriate site. Such an environment is provided in the thermonuclear
explosion of mass-accreting white dwarfs \cite{wall,howmey}. Exploratory, parameterized calculations for the canonical type Ia
supernova model -- the explosion of a white dwarf (WD) after it has accreted enough mass from a companion star to reach the
Chandrasekhar limit -- also found an underproduction of light p-nuclei as well as of $^{138}$La and $^{180m}$Ta, even when
assuming a seed enrichment of factors $10^3-10^4$ in s-process nuclei from an AGB companion \cite{arngorp}. The synthesis of
p-nuclei proceeds via a $\gamma$-process also in these types of models. Another alternative is a subclass of type Ia
supernovae which is supposed to be caused by the disruption of a sub-Chandrasekhar WD (mainly composed of C and O) due to a thermonuclear
runaway in a He-rich accretion layer \cite{gorsubchandra,travsubchandra}. High neutron fluxes are built up in the early phase
of the explosion and a weak r-process ensues. Once the temperature exceeds $T_9\approx2$ photodisintegrations take over and move the
nucleosynthesis to the proton-rich side where two processes act: the $\gamma$-process, as described above, and additionally
rapid proton captures on unstable nuclei at $3<T_9 \leq 3.5$. The latter is somewhat similar to the rp-process but at much lower proton densities and
thus closer to stability. The proton captures are hindered by ($\gamma$,p)-reactions but nuclei with low capture $Q$-value can be
bypassed by (n,p) reactions due to the available free neutrons \cite{arngorp,gorsubchandra}. This so-called \textit{pn-process} can efficiently
produce the light p-nuclides from Se to Ru but it overproduces them in relation to the heavier ones. Again, a strong increase
in the photodisintegration seed abundances would be required to produce all p-nuclei at solar relative abundances. (It has to be
noted that here the nuclei produced in the pn-process would be primary whereas the others are secondary.) Both WD scenarios
(canonical and sub-Chandrasekhar mass type Ia supernovae) suffer from the fact that they are difficult to simulate and
self-consistent hydrodynamical models including accretion, pre-explosive burning, explosion, and explosive burning in turbulent
layers are missing. A complete model would also allow to study the actual amount of seed enhancement (if any) in these
environments. Recent calculations based on trajectories from 2D simulations of WD explosions confirm the feasibility of
p-nucleus production in these sites \cite{travsubchandra}.

Considering even proton-richer conditions, two additional types of rapid proton capture processes may also yield p-nuclides from
the decay of very proton-rich progenitor nuclei. Both of them imply that p-production is primary.
The \textit{rp-process} occurs in explosive H- and He-burning on the surface of a
mass-accreting neutron star and involves sequences of proton captures and $\beta$-decays along the proton dripline \cite{schatz}.
How far the burning can move up the nuclear chart depends on details of the hydrodynamics (convection) and the amount of accreted
protons. There is a definitive endpoint, however, when the rp-process path runs into the region of Te $\alpha$-emitters \cite{endpoint}.
Therefore, if the rp-process actually runs that far, only p-nuclides with $A<110$ can be reached through decays of very proton-rich
progenitors. Currently it is also unclear whether the produced nuclides can be ejected into the interstellar medium or whether
they are trapped in the gravitational field and eventual only modify the surface composition of the neutron star. On the other hand,
very proton-rich conditions are found in the innermost ejected layers of a core-collapse supernova due to the interaction of
$\nu_\mathrm{e}$ with the ejected, hot, and dense matter at early times. The hot matter freezes out from nuclear statistical equilibrium
and sequences of proton captures and $\beta$-decays ensue, similar as in the late phase of the rp-process. The nucleosynthesis
timescale is given by the explosion and subsequent freeze-out and is shorter than in the rp-process. The flow towards heavier
elements would be hindered by nuclides with low proton capture $Q$-value and long $\beta$-decay halflives, similar as in the pn-process.
Again, (n,p) reactions accelerate the upward flow. The neutrons stem from the reaction $\bar{\nu}_\mathrm{e}+\mathrm{p}\rightarrow
\mathrm{n}+\mathrm{e}^+$. Although at early times the $\bar{\nu}_\mathrm{e}$ flux is small, the sheer number of free protons
guarantees a constant neutron flux. Thus, this process was termed the \textit{$\nu$p-process} \cite{frohlich,pruet}. Like the rp- and pn-processes
it may contribute to the light p-nuclides via decays of very proton-rich species. The details depend sensitively on the
explosion mechanism, the neutrino emission from the proto-neutron star, and the hydrodynamics governing the ejecta motion.

Historically, the longlived radionuclides $^{92}$Nb, $^{97}$Tc, $^{98}$Tc, $^{146}$Sm are not labelled as p-nuclides but they
cannot be produced in the s- and r-process, either. Measurements in meteoritic material firmly established the presence of
$^{92}$Nb and $^{146}$Sm at the formation of the solar system whereas only upper limits can be specified for the Tc isotopes \cite{dauphas}. The measured abundances of the above radionuclides are consistent within galactic chemical evolution models (i.e., they
yield a unique chemical evolution history).
The fact that $^{92}$Nb existed at the formation of the
solar system \cite{harper,schoenbaechler} puts tight constraints on the contributions of the rp-, np-, and $\nu$p-processes to the light p-nuclei.
The p-radionuclide $^{92}$Nb cannot be reached by decays from the proton-rich side because it is shielded by its stable isobar
$^{92}$Mo. Combining the abundance information on the stable p-nuclei with those of the p-radionuclides the
$^{92}$Nb/$^{92}$Mo production ratio $r_\mathrm{NbMo}$ in the Galaxy
can be inferred: $1.7\pm 0.5\times 10^{-3} < r_\mathrm{NbMo} \leq 0.1$ \cite{dauphas}. The $^{92}$Mo production in the $\gamma$-process
is too low by a factor of about 10. If the remaining 90\% stem from decays of very proton-rich nuclei (which cannot produce
$^{92}$Nb) this would yield a production ratio of $1.5\pm 1.0\times 10^{-4}$ which is clearly outside the permitted range for
$r_\mathrm{NbMo}$. This puts severe doubts on any significant contribution of the rapid proton capture processes to nuclei with
$A\geq 92$.
(A similar argument applies to the production of $^{98}$Tc but it is at higher mass and its early solar system abundance is not well
known.)

\section{Nuclear physics uncertainties}

\noindent
Despite of the number of suggested sites, the nuclear physics underlying the p-production is similar in all of them, especially if
a contribution of rapid proton capture processes far from stability can be ruled out. The nuclei involved in the $\gamma$-process
are the stable nuclides and moderately unstable, proton-rich nuclei. The masses have been measured and thus the reaction $Q$-values
are well known. Halflives are known in principle but the electron captures and $\beta^+$-decays need to be modified in the stellar plasma
by the application of theoretical corrections for ionization and thermal excitation.

Stellar correction factors act strongly also in the photodisintegration reactions (and captures). At the relatively high
temperatures encountered in p-nucleosynthesis, a considerable fraction of nuclei is thermally excited. Additional reactions
proceeding from the excited states of nuclei become possible. Thus, laboratory measurements always only include
a fraction of the astrophysically relevant transitions. To minimize the correction effects, reactions should always be
determined in the direction of positive $Q$-value \cite{raureview}. It can be shown, however, that also charged particle captures with negative
$Q$-values show smaller corrections than their photodisintegration counterparts, due to the Coulomb suppression of the
stellar enhancement \cite{tomsupp}. Similar considerations apply for (p,n) and (n,p) reactions \cite{kisssupp}. Unfortunately,
measurements of the (p,$\gamma$) and ($\alpha$,$\gamma$) cross sections close to the astrophysically relevant energy windows of several MeV \cite{windows} at
$2\leq T_9 \leq 3$ (or $3 \leq T_9 \leq 3.5$ for proton captures in the pn- and $\nu$p-processes)
are scarce even for stable nuclei \cite{zsoltthis,ozkanthis}. Therefore the full $\gamma$-process
reaction network includes a majority of rates based on theoretical predictions. Required are (n,$\gamma$), (p,$\gamma$), and ($\alpha$,$\gamma$) reactions and their inverses on stable nuclei and up to about 15 units into the region of proton-rich, unstable nuclei
(proton-richer nuclei are required in the rapid proton capture processes \cite{weber}). Not only the pn- and $\nu$p-processes contain (n,p) reactions, they are also important in the $\gamma$-process where they
are initiated by the neutrons released through ($\gamma$,n) and speed up the matter flow back to stability \cite{rapp}. It should
be kept in mind that the theoretical corrections for the thermally excited nuclei have to be predicted even if a full database of
experimental (ground state) rates were available.

Charged particle captures and photodisintegrations are only sensitive to the optical potential at the astrophysically
relevant low energies. These optical potentials are usually derived from elastic scattering at higher energies and are
not well constrained around the Coulomb barrier. Especially the imaginary part should be energy-dependent (see \cite{raureview} for more details on optical potentials in astrophysical rate predictions and the influence of other nuclear properties).
Comparison of theoretical predictions with the few available data at low energy revealed a systematic 2-3 times overprediction of the ($\alpha$,$\gamma$) cross sections when using the ``standard'' potential of \cite{mcfadden} (see, e.g., \cite{zsoltthis,ozkanthis,tm169,yalcin,gyurky,151eu} and references therein). So far, the
only known example of a larger deviation was found in $^{144}$Sm($\alpha$,$\gamma$)$^{148}$Gd (determining the $^{144}$Sm/$^{146}$Sm production ratio in the $\gamma$-process) where the measured cross section is lower than the standard prediction by more than
an order of magnitude at astrophysical energies \cite{sm144}. Since this conclusion hinges only on a few datapoints with comparatively large errors, an independent (re-)measurement of this reaction is important.

While the standard $\alpha$+nucleus potential is a purely phenomenological one, the optical potentials for protons and neutrons used
in the prediction of astrophysical rates and also the interpretation of nuclear data are based on a more microscopic treatment,
a Br\"uckner-Hartree-Fock calculation with the Local Density Approximation including nuclear matter density distributions which are
in turn derived from microscopic calculations \cite{raureview,jlmm,bauge1,bauge2}. Nevertheless, they also may show some
deficiencies at low energies because also such descriptions use scattering data to constrain their free parameters \cite{vector}. In comparison
to measured low energy (p,$\gamma$) cross sections, these potentials fare much better than the one for the $\alpha$+nucleus optical potential, being often in very good agreement with the data, sometimes being off by a maximal factor of two (see, e.g., \cite{raureview,pdpg,cdpg} and references therein). In general, the uncertainty in the astrophysical rates caused by the nucleon optical potentials seems to be much lower than for the $\alpha$-capture rates. Recent (p,$\gamma$) and (p,n) data of higher precision close to the astrophysically relevant energy window, however, revealed a possible need for modification of the imaginary part of the nucleon potentials \cite{raureview,cdpg,npa4proc}.
A consistently increased strength of the imaginary part at low energies improves the reproduction of the experimental data for a number of
reactions.

\section{Deflections and branchings}

\begin{figure}
\centerline{\includegraphics[width=0.9\textwidth]{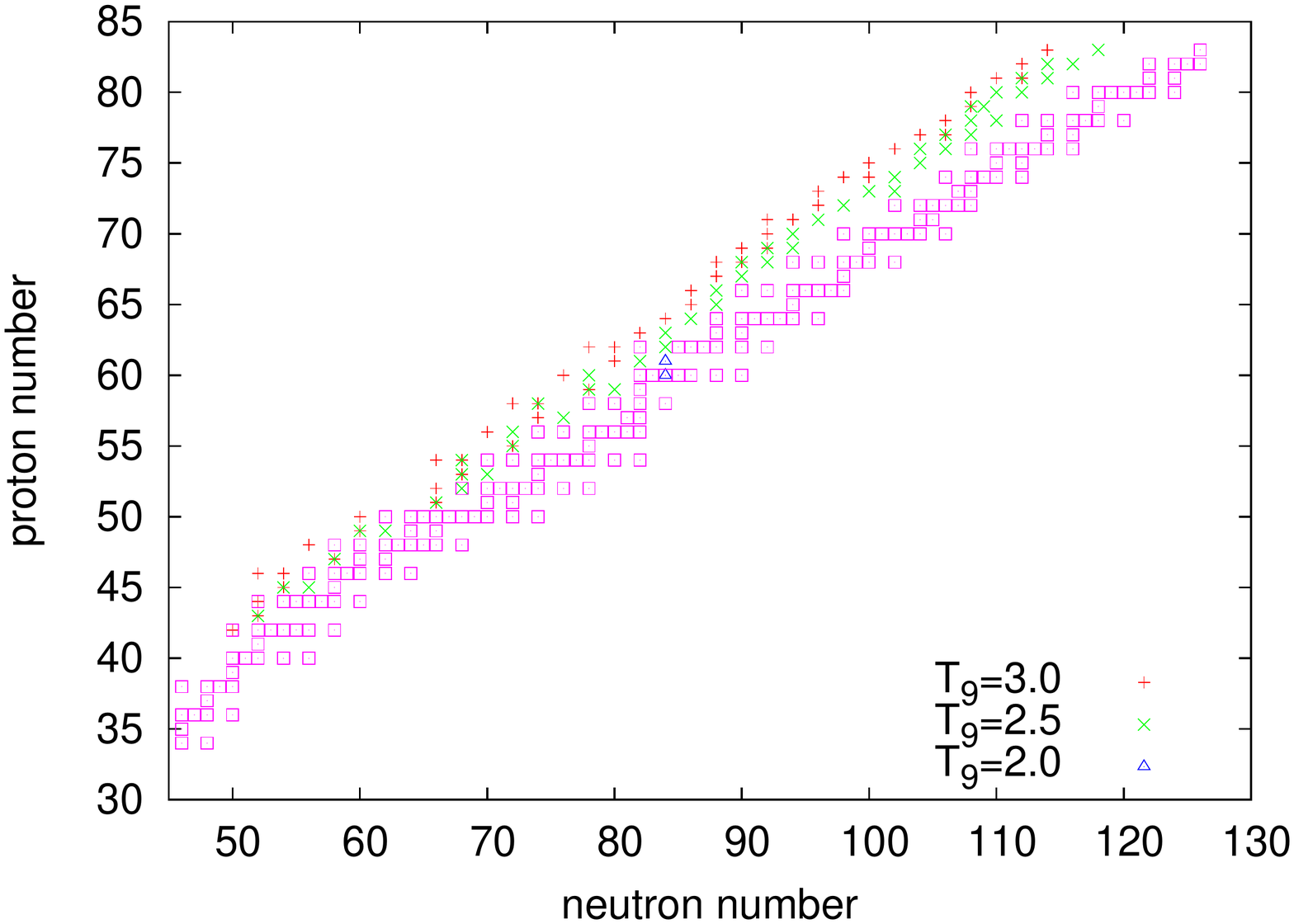}}
\caption{Deflection points in the photodisintegration path for three stellar temperatures; natural isotopes are marked by squares.
At low temperature, often the ($\gamma$,n) reactions cease within an isotopic chain before reaching a path deflection.}
\label{fig:deflect}
\end{figure}

\noindent
What is the actual impact of uncertainties in the photodisintegration rates on the production of the p-nuclides? To answer this
question -- which is also relevant for determining the main targets of future investigations -- it has to be realized that not
all reactions are equally important in all sections of the nucleosynthesis network. Furthermore, their relative importance changes
with the plasma temperature reached in the respective mass zone where p-nuclei are produced. In a $\gamma$-process,
light p-nuclei are predominantly produced at higher temperatures (allowing efficient photodisintegration of the nuclei around mass $A\approx 100$)
whereas the production maximum of the heavy species lies towards the lower end of the temperature range \cite{arngorp,rapp}. Neutron captures and especially
($\gamma$,n) reactions are important throughout the $\gamma$-process network as the photodisintegration of stable nuclides
commences with ($\gamma$,n) reactions until sufficiently proton-rich nuclei have been produced and ($\gamma$,p) or ($\gamma$,$\alpha$)
reactions become faster. The released neutrons can be captured again by other nuclei and push the reaction path back to stability in
the region of the light p-nuclides.

Studies of the $\gamma$-process with systematic variations of the ($\gamma$,p) and ($\gamma$,$\alpha$) rates by constant factors
\cite{rapp} or using a selection of different optical potentials for the ($\gamma$,$\alpha$) rates \cite{branchings} confirm
that uncertainties in the ($\gamma$,p) rates mainly affect the lower half, whereas those in the ($\gamma$,$\alpha$)
rates affect the upper half of the p-nuclide mass range. This can be easily understood by examination of the reaction paths and
the competition between the photodisintegration reactions. In each isotopic chain we commence with initial ($\gamma$,n) on stable
isotopes and move towards the proton-rich side. The proton-richer a nucleus, the slower the ($\gamma$,n) rate while
($\gamma$,p) and/or ($\gamma$,$\alpha$) rates increase. At a certain isotope within the isotopic chain, a charged particle emission
rate will become faster than ($\gamma$,n) and thus deflect the reaction sequence to lower charge number. Historically, this endpoint
of the ($\gamma$,n) chain has been called ``branching'', inspired by the branchings in the s-process path \cite{woohow,branchings}.
A more appropriate term would be ``deflection (point)'', though, because -- unlike in the s-process -- the reaction path does not split
into two branches. The relative changes of the photodisintegration rates from one isotope to the next are so large that at each
isotope only one of the emissions dominates. There are very few cases of two (or even three) types of emissions being comparable
and those strongly depend on the optical potentials used. Examination of the deflection points easily shows that at higher mass we
encounter ($\gamma$,$\alpha$) deflections whereas at lower mass most deflections are caused by ($\gamma$,p) due to the
distribution of reaction $Q$-values and Coulomb barriers \cite{branchings}.

Figure~\ref{fig:deflect} shows the $\gamma$-process deflection points for three relevant temperatures. The calculations have been
performed with the SMARAGD code \cite{raureview,smaragd} using the standard optical potentials of \cite{jlmm} and \cite{mcfadden}.
They supersede the results obtained in \cite{branchings}.
It is obvious that the higher the plasma temperature the further into the proton-rich side ($\gamma$,n) reactions can act. At low
temperature, even the ($\gamma$,n) reaction sequence may not be able to move much beyond the stable isotopes because it becomes
too slow compared to the explosive timescale (and neutron capture may be faster).
Obviously, ($\gamma$,p) and ($\gamma$,$\alpha$) are even slower in those cases and the photodisintegrations are ``stuck'', just slightly reordering stable abundances. For the plots presented here, photodisintegration rates slower than 0.2 s$^{-1}$ were neglected.

\begin{figure}
\centerline{\includegraphics[width=0.9\textwidth]{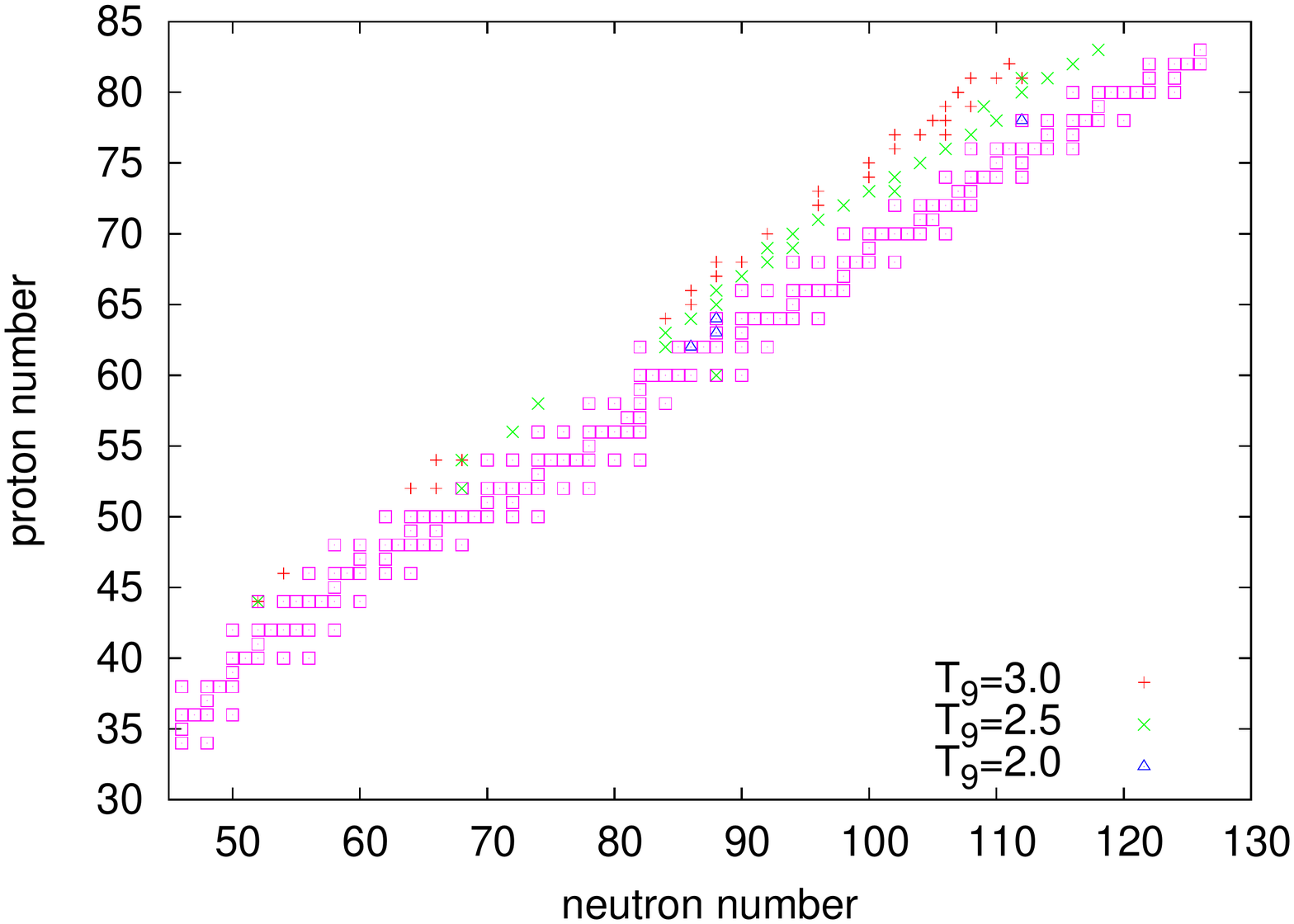}}
\caption{Changes in the deflections when dividing the ($\gamma$,$\alpha$) rate by a factor of 10. Only nuclei are marked which
show different deflections than the standard case shown in Fig.~\protect\ref{fig:deflect}.}
\label{fig:uncertainalphadeflect}
\end{figure}

The above said, it has become obvious that not all possible reactions in the $\gamma$-process network have to be known with high
accuracy. Rather, only the dominant reaction sequence has to be known accurately, and within this sequence the slowest
reaction as it determines the amount of processing within the given short timescale of the explosive process. A large error
in a rate which is slower by several orders of magnitude than the dominating rate is inconsequential. Charged particle rates are
important only at or close to the deflection points because they are many orders of magnitude smaller than the ($\gamma$,n) rates
on neutron-richer isotopes. The impact of suppressing all ($\gamma$,$\alpha$) rates by a factor of ten is shown in Fig.~\ref{fig:uncertainalphadeflect}. Only a comparatively small number of deflections are altered. Despite of the result of
\cite{sm144}, such a large change may be exaggerated. Other reaction data suggested suppressions by factors of 2-3 (see previous section).
When recalculating the deflections using the potential of \cite{frohdip,raufroh} (which provides a good reproduction of the
low-energy reaction data), there are even fewer modifications. Compared to what is shown in Fig.~\ref{fig:deflect} an additional ($\gamma$,n)/(($\gamma$,p)+($\gamma$,$\alpha$)) branching at $^{187}$Ir appears at $T_9=2.5$, and a
($\gamma$,p)/(($\gamma$,n)+($\gamma$,$\alpha$)) branching at $^{100}$Pd is replaced by a ($\gamma$,p) deflection at $T_9=3.0$,
thereby eliminating the ($\gamma$,p) deflection at $^{98}$Pd.

\begin{figure}
\centerline{\includegraphics[width=\textwidth]{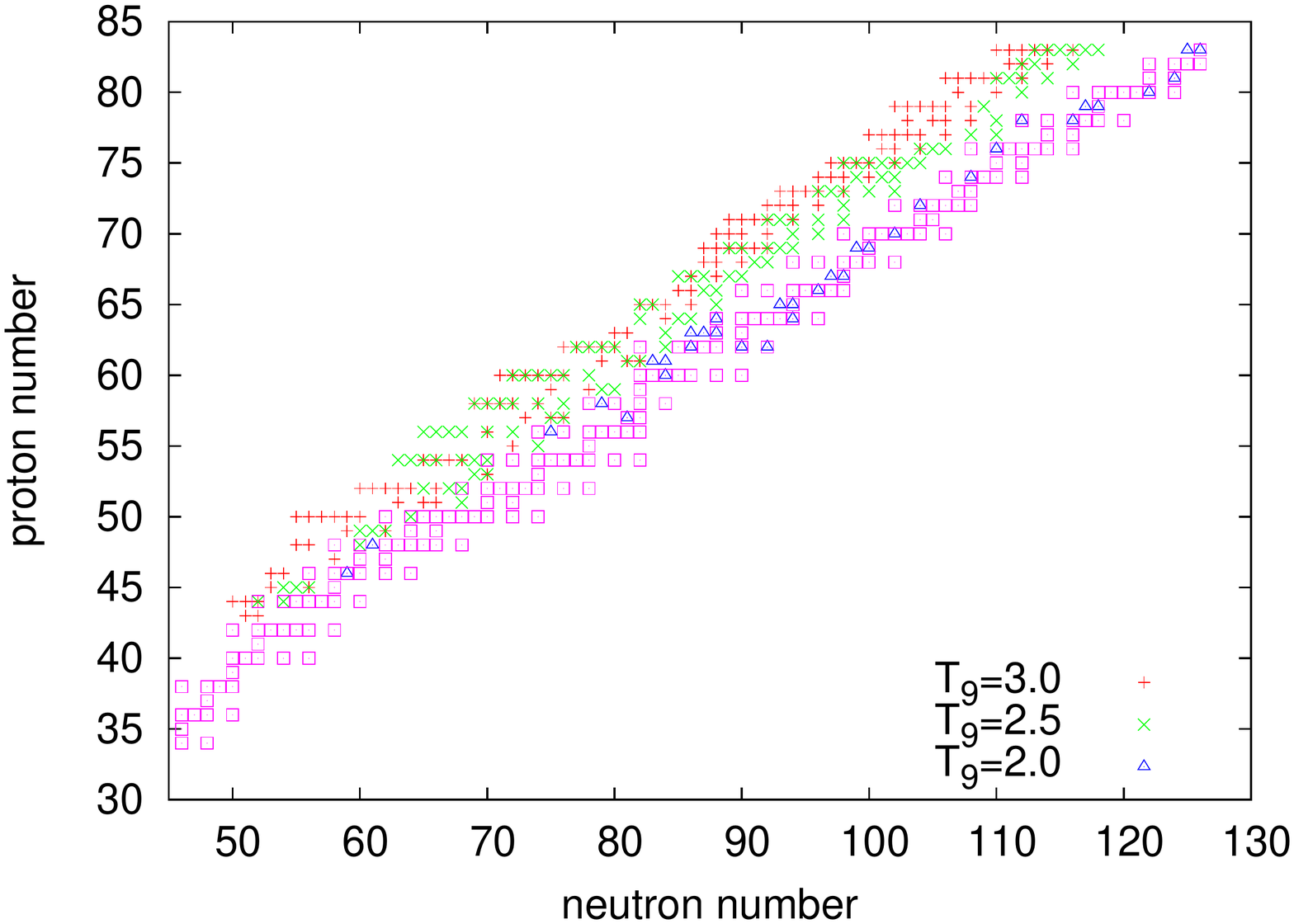}}
\caption{Uncertain deflections and branchings when simultaneously varying ($\gamma$,n), ($\gamma$,p), and ($\gamma$,$\alpha$) rates.}
\label{fig:uncertain}
\end{figure}

A simultaneous variation of all photodisintegration rates within reasonably chosen limits, encompassing the maximally expected
rate uncertainty, can be taken as the extreme case. Figure
\ref{fig:uncertain} shows the changed deflections (and branchings) when simultaneously varying the ($\gamma$,n) rate by factors 1, 2, and 0.5, the ($\gamma$,p) rate by factors 1, 2, and $1/3$, and the ($\gamma$,$\alpha$) rate by factors 1 and 0.1 (for each nucleus
separately). Again, the ($\gamma$,n) photodisintegrations cannot efficiently destroy the stable isotopes of some elements at low
temperature. 

\section{Conclusion}

\noindent
We have come a long way since \cite{b2fh,agw} but the mystery of the origin of the p-nuclides is still with us.
Although there are considerable uncertainties in the astrophysical (hydrodynamical) modeling of the sites possibly producing
p-nuclei, a sound base of nuclear reaction rates is essential for all such investigations. As long as an experimental determination
of the rates around the deflection points is impossible, measurements of low energy cross sections of stable nuclides
are essential to test
and improve the theoretical calculations. This has been underlined by the recent results regarding optical potentials for the
interaction of charged nuclei. Further measurements (at even lower energy) are highly desireable and, along with
improved, self-consistent hydrodynamic simulations of the possible production sites, will gradually improve
our understanding of p-nucleosynthesis.

\end{document}